\documentclass[12pt,A4]{article}
\usepackage{epsfig}

\begin{document}
\begin{center}
{\bf\large Polarization effects in the quasi - elastic {\it (p,2p)}\\
reaction
with the nuclear {\it S} - shell protons at 1 GeV }\\
\end{center}

\begin{center}
{\bf O.~V.~Miklukho, A.~Yu.~Kisselev, D.~A.~Aksenov, G.~M.~Amalsky,
V.~A.~Andreev, S.~V.~Evstiukhin, O.~Ya.~Fedorov, G.~E.~Gavrilov,
A.~A.~Izotov, L.~M.~Kochenda, M.~P.~Levchenko, D.~A.~Maysuzenko,
V.~A.~Murzin, D.~V.~Novinsky, A.~N.~Prokofiev, A.~V.~Shvedchikov,
V.~Yu.~Trautman, S.~I.~Trush, A.~A.~Zhdanov}\\
\end{center}

\begin{center}
{\it B.P.Konstantinov Petersburg Nuclear Physics Institute,
Gatchina, Russia}\\
\end{center}

The polarization of the secondary protons  in the
$(p,2p)$ reaction with the $S$-shell protons of nuclei $^4$He, $^6$Li,
$^{12}$C, $^{28}$Si, $^{40}$Ca was measured at 1 GeV unpolarized
proton beam. The spin correlation parameters $C_{ij}$ for the
$^4$He and $^{12}$C targets also were for the first time obtained as well.
The polarization measurements were performed by means of a two-arm
magnetic spectrometer, each arm of which was equipped with the
multiwire-proportional chamber polarimeter. This experiment
was aimed to study a modification of the proton-proton
scattering matrix in the nuclear medium.\\

 {\bf Comments:} 28 pages, 7 figures, 6 tables, Submitted to the journal "Physics of Atomic Nuclei"\\

 {\bf Category:} Nuclear experiment (nucl-ex)\\
\newpage

\section{Introduction}
~~~~There were some speculations on  modifications of nucleon and meson masses
and sizes, and of meson-nucleon coupling constants, and, as
consequence, of a nucleon-nucleon scattering matrix in nuclear medium [1-6].
These speculations were
motivated by a variety of theoretical points of view, including the
renormalization effects due to strong relativistic nuclear fields,
deconfinement of quarks, and partial chiral symmetry restoration.

 This  work is a part of the experimental program in the frame of which
 the medium-induced modifications of the nucleon-nucleon scattering amplitudes are
 studied at PNPI synchrocyclotron with 1 GeV proton beam [7-13].
The intermediate-energy quasi-free (p,2p) reaction is a good experimental
way to study such effects, since in the first approximation this reaction can
be considered as a proton-proton scattering in the nuclear matter. Usage of
S-shell protons (with zero orbital momentum) is prefered because interpretation
of obtained data in this case is essentially simplified since the
effective polarization is not involved \cite{Jacob1976}.
The polarization observables in the reaction are compared with
those in the  elastic $pp$ scattering.
In our exclusive experiment a two-arm magnetic spectrometer
  is used, the shell structure of
 the nuclei being evidently distinguished. To measure polarization characteristics of the
 reaction, each arm of the spectrometer was equipped with the multiwire-proportional chamber
 polarimeter.

 In early PNPI-RCNP experiment \cite{Andreev2004}, the polarizations $P_1$  and $P_2$ of both
 secondary protons from the $(p,2p)$
 reactions at 1 GeV with the 1$S$-shell protons of the nuclei $^6$Li, $^{12}$C and
 with the 2$S$-shell protons of the $^{40}$Ca nucleus has been measured at
 the nuclear proton momenta
 close to zero. The polarization observed in the experiment,
 as well as the analyzing power $A_y$ in the RCNP experiment
at the 392 MeV polarized proton beam \cite{Hatanaka1997, Noro2000},
 drastically differed from that calculated in the framework of non-relativistic Plane Wave Impulse
 Approximation (PWIA) and of spin-dependent Distorted Wave Impulse Approximation (DWIA) \cite{Chant1983},
based on free space proton-proton interaction. This difference was found to have a negative value
and increases monotonously with the effective mean nuclear density
$\bar\rho$ \cite{Hatanaka1997}. The latter  is determined by an
 absorption of initial and secondary protons in nucleus matter.
 The observed inessential difference between the non-relativistic PWIA and DWIA
 calculations pointed out only to a small
depolarization of the secondary protons because of  proton-nucleon rescatterings inside a nucleus.
 All these facts strongly indicated
a modification of the proton-proton scattering amplitudes due to the modification of the main properties
of hadrons in the nuclear matter.

 Later, the result of the experiment with the $^4$He target
broke the mentioned above dependence of a difference
between the experimental polarization values  and those
calculated in the framework of the PWIA on the effective mean nuclear
density $\bar\rho$ \cite{Miklukho2006}. The difference for the $^4$He nucleus proved
to be close to that for the $^6$Li nucleus. This evidently contradicts to the elastic
proton-nucleus scattering experiment. According to the experiment, the $^4$He nucleus
has the largest mean nuclear density.
On the other hand, the  mentioned above deviation from the PWIA keeps to
be a monotonous function of the $S$-shell proton binding energy
$E_s$ for all investigated nuclei. It is possible that in the
light nuclei at least (with atomic number A$< $12), where the nuclear matter
is strongly heterogeneous, the
value of $\bar\rho$ does not properly reflect the scale of the
nuclear medium influence on the properties of the nucleon-nucleon interaction,
and the value of $E_s$ may also be a measure of this influence.
The important feature of the experiment with the $^4$He nucleus
was a possibility to see the medium effect without any contribution from the multi-step processes
(for instance, from the  $(p,2pN)$ reactions). These processes  could take place when
there were nucleons of outer shells as in other nuclei. Therefore they
could not also be an origin of the systematic difference between the polarizations
$P_1$ and $P_2$  clearly obtained for the first time in the experiment
\cite{Miklukho2006}.

Here we present the polarization data for the reaction with the nuclei
$^4$He, $^6$Li, $^{12}$C (1$S$-shell) and $^{40}$Ca nucleus (2$S$-shell).
These data were obtained by averaging the results of our early experiments [10,11]
and the results of the new measurements
performed after upgrade of the spectrometer electronics.
In the experiments the polarization was measured with much better statistical accuracy
for the nuclei $^4$He, $^{12}$C.
We also present new data on the polarization
in the reaction with the 1$S$-shell protons
of the $^{28}$Si nucleus. The 1$S$-state of the  $^{28}$Si nucleus has
the larger value of the mean proton binding energy $E_s$ (50 MeV) in comparison
with  that of the $^{12}$C nucleus (35 MeV).

Due to the new fast electronics CROS-3  of the spectrometer \cite{Bondar2007},
the experimental program was extended to measure the spin
correlation parameters $C_{ij}$  in the $(p,2p)$ reaction with nuclei.
The useful counting rate of the reaction drastically drops with increasing the
nucleus atomic number due to an absorption of the initial and secondary protons
in the nucleus matter. We measured these parameters in the reaction with
the light nuclei $^4$He and $^{12}$C.
The main interest was concentrated on measuring the spin correlation
parameter $C_{nn}$  since its value is the same in the
center-of-mass and laboratory systems. Besides, this parameter is not distorted by
the magnetic fields of the two-arm spectrometer because of the proton
anomalous magnetic moment \cite{Lock1973}. Since the polarization  and  the
spin correlation parameter $C_{nn}$ are expressed differently  through the
scattering matrix elements \cite{Krein1995}, the measurement of both these
polarization observables  can
give more comprehensive information about a modification of the hadron
properties in the nuclear medium.

\section{Experimental method}
~~~ The general layout of the
experimental setup is shown in Fig.~1.
\begin{figure}
\centering\epsfig{file=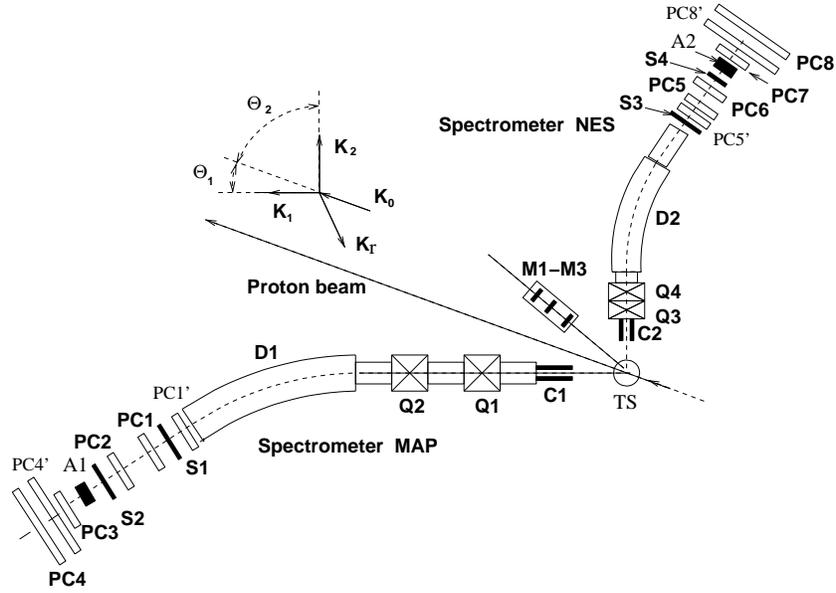, width=.8\textwidth}
\caption{\small The experimental setup. TS is the target of two-arm spectrometer;
Q1$\div$Q4 are the magnetic quadrupoles; D1, D2 are the dipole
magnets; C1, C2 are the collimators; S1$\div$S4 and M1$\div$M3 are
the scintillation counters; PC1$\div$PC4, PC1', PC4'
(PC5$\div$PC8, PC5', PC8') and A1 (A2) are the proportional
chambers and the carbon analyzer of the high-momentum (low-momentum)
polarimeter, respectively. The kinematics for the $(p,2p)$ reaction is
shown.}
\end{figure}
The experiment is performed at non-symmetric scattering angles
of the final state protons in the coplanar quasi-free scattering
geometry with a complete reconstruction of the reaction
kinematics. The measured secondary proton momenta $K_1$, $K_2$ (kinetic energies $T_1$, $T_2$)
and the scattering angles $\Theta_1$, $\Theta_2$ are used together
with the proton beam energy $T_0$ to calculate nuclear proton
separation energy $\Delta E$=$T_0$-$T_1$-$T_2$ and the residual nucleus momentum {\bf $K_r$}
for each $(p,2p)$ event. In the impulse approximation, the $K_r$
is equal to the momentum $K$ of the nuclear proton before the
interaction (${\bf K_r}$=-${\bf K}$).

 External proton   beam of the PNPI synchrocyclotron was focused
onto the  target TS of a two-arm spectrometer consisted of the magnetic
spectrometers MAP and NES. The beam intensity was monitored by
the scintillation telescope M1, M2, M3 and was about of
5$\cdot$10$^{10}$ protons/(s$\cdot$cm$^2$).

The solid nuclear targets TS made from CH$_2$ (for the setup
calibration), $^6$Li, $^{12}$C, $^{28}$Si, $^{40}$Ca (Table~1) and
the universal cryogenic target with the liquid helium  $^4$He (or
with the liquid hydrogen for calibration) were used in the
experiment \cite{Miklukho2006,Kotchenda2009}.

\begin{table}
\caption{Solid target parameters}
\label{table:targets}
\begin{center}
\begin{tabular}{l|c|c}
\hline
Target               &  Dimensions (mm)           & Isotope concentration ($\%$) \\
\hline
                     &  diameter x length           &\\
CH$_2$               &  22x70                 &         \\
                  & thickness x width x height           & \\
$^6$Li                  & 4.5x12x25             &   99.0 \\
$^{12}$C                & 4.0x18x70            &   98.9 \\
$^{28}$Si               & 6.0x25x70            &   99.9 \\
$^{40}$Ca               & 4.0x10x13            &   97.0 \\
\hline
\end{tabular}
\end{center}
\end{table}
%\newpage
 Cylindrical aluminium appendix of the
cryogenic target had the dimensions: diameter - 65 mm,
length - 70 mm, wall thickness - 0.1 mm. The diameter of the beam
spot on the target was less than 15 mm.
\begin{table}
\caption{Parameters of the magnetic spectrometers}
\label{table:spectrometers}
\begin{center}
\begin{tabular}{l|c|c}
\hline
Spectrometer                                   & NES                & MAP\\
\hline
Maximum particle momentum [GeV/c]              & 1.0                & 1.7\\
Axial trajectory radius $\rho$ [m]             & 3.27               & 5.5\\
Deflection angle $\beta$, [deg]                & 37.2               & 24.0\\
Dispersion in the focal plane $D_f$, [mm/$\%$] & 24                 & 22\\
Solid angle acceptance $\Omega$, [sr]          & $3.1\cdot 10^{-3}$ & $4.0\cdot 10^{-4}$\\
Momentum acceptance $\Delta K/K$, [$\%$]       & 8.0                & 8.0\\
Energy resolution (FWHM), [MeV]                & $\sim 2.0$         & $\sim 1.5$\\
\hline
\end{tabular}
\end{center}
\end{table}
The spectrometers were used for registration of the
secondary protons from the $(p,2p)$ reaction in coincidence and for
measurement of their momenta and outgoing angles. The polarization
of these protons $P_1$ and $P_2$, and the spin correlation
parameters $C_{ij}$ were measured by the polarimeters located in
the region of focal planes of the spectrometers MAP and NES (Fig.~1).
The left index of the $C_{ij}$,
$i$ ($i$ is $n$ or $s^,$), and the right index $j$ ($j$ is $n$ or $s^{,,}$)
correspond to the forward scattered proton analysed by the
MAP polarimeter and the recoil proton analysed by the NES
polarimeter, respectively. The unit vector ${\bf n}$ is
perpendicular to the  scattering plane of the  reaction.
Unit vectors ${\bf {s^,}}$ and ${\bf s^{,,}}$
are perpendicular to the vector ${\bf n}$ and, the coordinate axises
z$^,$ and z$^{,,}$ (Fig.~1) of the polarimeters.

The main parameters of the two-arm magnetic spectrometer and
polarimeters are listed in Table~2 and Table~3, respectively.
\newpage
\begin{table}
\caption{Polarimeter parameters}
\label{table:polarimeters}
\begin{center}
\begin{tabular}{l|c|c}
\hline
Polarimeter                    & NES           & MAP\\
\hline
Carbon block thickness [mm]    & $79$          & $199$\\
Polar angular range [deg]      & $6 \div 18$ & $3 \div 16$\\
Average analyzing power        & $\geq 0.46$   & $\geq 0.23$\\
Efficiency [$\%$]              & $\sim 2$      & $\sim 5$\\
\hline
\end{tabular}
\end{center}
\end{table}
The
$\Delta E$ resolution of the spectrometer estimated from the elastic
proton-proton scattering (Fig.~2, the left panel) with the 22-mm-thick cylindrical CH$_2$
target (see Table~1) was found to be about of 5 MeV (FWHM).
\begin{figure}
\centering\epsfig{file=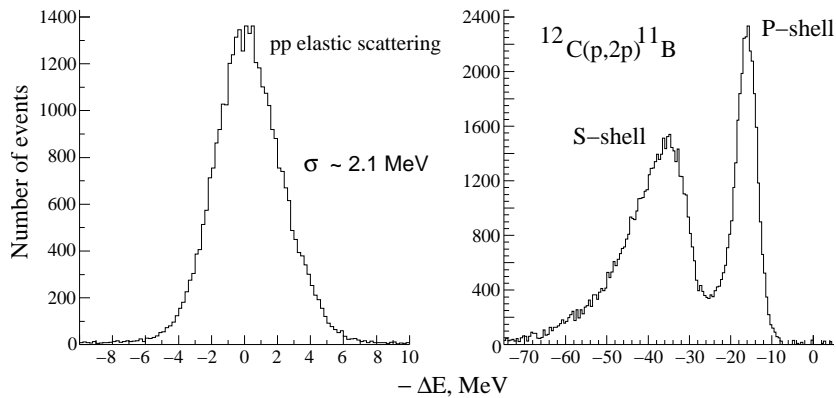, width=.8\textwidth}
\caption{\small Proton separation energy spectra for the elastic $pp$ scattering (the left panel) and
       for the $(p,2p)$ reaction with the $^{12}$C nucleus (the right panel). In the $^{12}$C spectrum,
       the contribution from the accidental coincidence background was subtracted.}
\end{figure}

The track information from the proportional chambers of both
polarimeters was used in the offline analysis to find the
azimuthal $\phi_1$, $\phi_2$ and polar $\theta_1$, $\theta_2$
angles of the proton scattering from the analyzers A1, A2 for each
$(p,2p)$ event. In the case of absence of the accidental coincidence
background (the case of the elastic proton-proton scattering)
the polarization parameters could be found as \cite{Fedorov2001}
\begin{equation}
\label{equation:p12}
P_{1,2}=\frac{ 2<\cos\phi_{1,2}> }{ <A(\theta_{1,2},K_{1,2})> }\\,
\end{equation}
\begin{equation}
\label{equation:cnn}
C_{nn}=\frac{  4<\cos\phi_1\cos\phi_2> }{<A(\theta_1,K_1)><A(\theta_2,K_2)> }\\,
\end{equation}
\begin{equation}
\label{equation:cs's''}
C_{s^,{s^{,,}}}=\frac{  4<\sin\phi_1\sin\phi_2> }{<A(\theta_1,K_1)><A(\theta_2,K_2)> }\\,
\end{equation}
\begin{equation}
\label{equation:cns''}
C_{n{s^{,,}}}=\frac{  4<\cos\phi_1\sin\phi_2> }{<A(\theta_1,K_1)><A(\theta_2,K_2)> }\\,
\end{equation}
\begin{equation}
\label{equation:cs'n}
C_{s^,{n}}=\frac{  4<\sin\phi_1\cos\phi_2> }{<A(\theta_1,K_1)><A(\theta_2,K_2)> }\\,
\end{equation}
where averaging was made over a set of events within the working
angular range of $\theta_{1,2}$ (see Table~3) for the
polarimeters. $A(\theta_1,K_1)$ and $A(\theta_2,K_2)$, which were
averaged over the same set of events, are the carbon analyzing
power parameterized according to \cite{Fedorov1979} and \cite{Waters1978} for the MAP and NES
polarimeter, respectively.

We estimated the polarization parameters by
folding the theoretical functional shape of the azimuthal angular
distribution into experimental one \cite{Miklukho2010}, using the CERNLIB MINUIT
package \cite{James1998} and likelihood $\chi^2$ estimator \cite{Baker1984}. This method
permits to realize the control over $\chi^2$ in the case the
experimentally measured azimuthal distribution is distorted due to
the instrumental problems.

The time difference (TOF) between the signals from the
scintillation counters S2 and S4, was measured. It
served to  control  the accidental coincidence background. The
events from four neighboring proton beam bunches were recorded.
Three of them contained the background events only and were used in
the offline analysis to estimate the background polarization
parameters and the background contribution at the main bunch containing
the correlated events \cite{Kisselev2012}. The background
polarization $P_1$ was found to be about 0.20-0.25  for all investigated nuclei.
The polarization $P_2$ was less than 0.08. This value corresponds
to the $^4$He nucleus.

  The recoil proton spectrometer NES  was installed at
a fixed angle $\Theta_2\simeq$ 53.2$^\circ$.
At a given value of the $S$-shell mean binding energy of nucleus
under investigation, the angular and momentum settings of the MAP
spectrometer and the momentum setting of the NES spectrometer were
chosen to get a kinematics of the $(p,2p)$ reaction close to that of the
elastic proton-proton scattering. In this kinematics,
momentum  $K$ of the nuclear proton before the interaction is
close to zero. At this condition, the counting rate of the $S$-shell proton
knockout reaction should be maximal. In Fig.~2 (the right panel) and Fig.~3, the proton
separation energy spectra for the  reaction with the $^{12}$C and
$^{28}$Si nuclei are shown.
These spectra were obtained by subtracting the accidental coincidence background.
\begin{figure}
\centering\epsfig{file=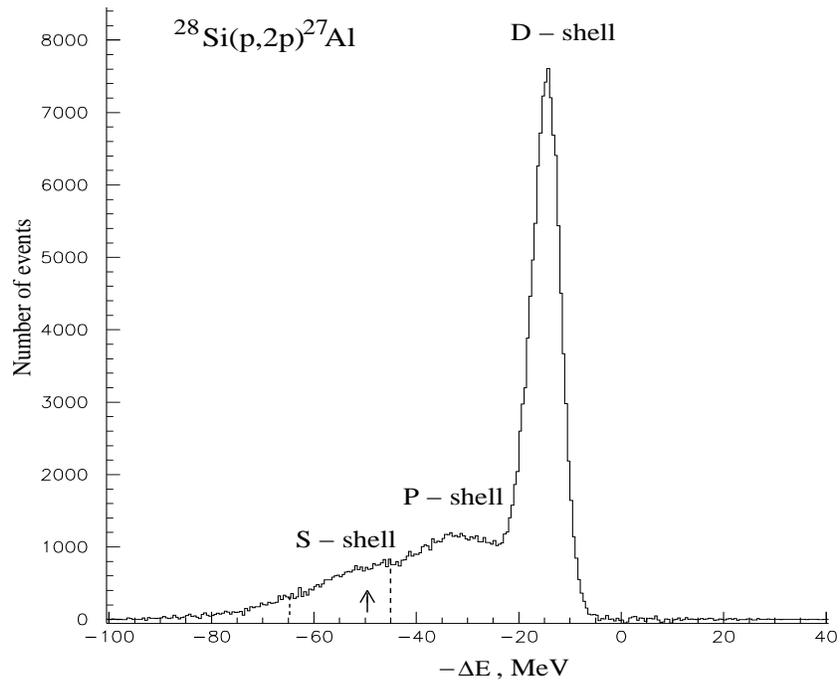,width=.8\textwidth,height=90mm}
\caption{\small Proton separation energy spectrum for the reaction $^{28}$Si$(p,2p)^{27}$Al.
The contribution from the accidental coincidence background was subtracted.
The area between the vertical dashed lines was used for data analysis.}
\end{figure}
As seen in Fig.~3, even at the preferable condition for the $S$-shell
proton knockout process, the contribution from the scattering off
the external shell protons for the  heavy nucleus $^{28}$Si is dominant.
The number of the subtracted background events in the figure  was 
approximately equal to the correlated
ones in the $\Delta E$ region (between the dashed lines) used in offline analysis.

Measurements of the spin correlation parameters and even of the
polarization in the $(p,2p)$ reaction with heavy nuclei became
possible due to the fast proportional chamber readout system,
 developed and produced at PNPI \cite{Bondar2007}. This electronics
allowed us to collect the correlation events without distortion at a
high rate of the accidental coincidence background.

\section{Results and discussion}
~~~ In Fig.~4 and Table~4 in Appendix, the polarizations $P_1$, $P_2$  in the $(p,2p)$ reaction
with the $S$-shell protons of the nuclei $^4$He, $^6$Li, $^{12}$C, $^{28}$Si, $^{40}$Ca
are plotted versus  the $S$-shell proton binding energy $E_s$.
\begin{figure}
\centering\epsfig{file=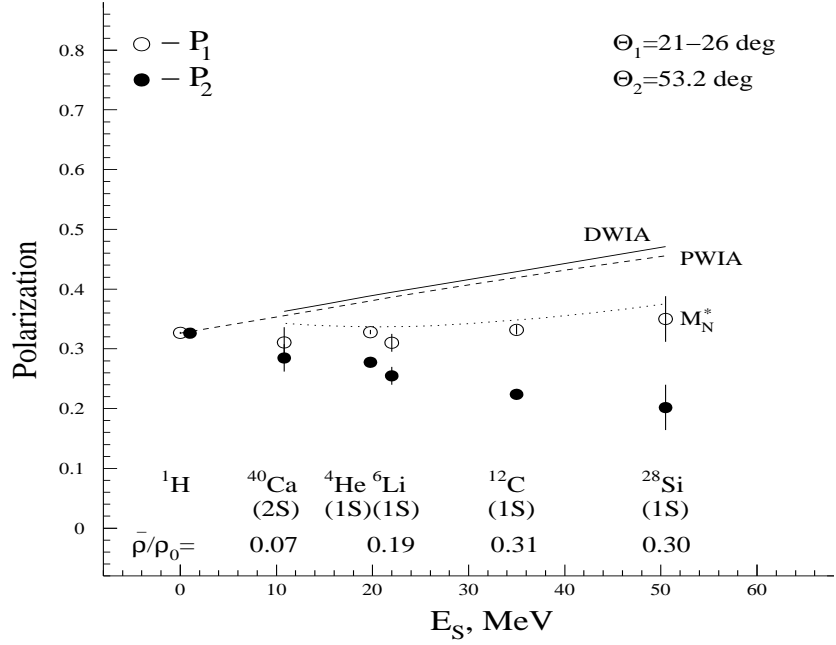,width=.8\textwidth,height=85mm}
\caption{\small Polarizations $P_1$ and $P_2$  of the  protons scattered
at the angles $\Theta_1$  ($\circ$) and $\Theta_2\simeq$ 53.2$^\circ$
($\bullet$) in the $(p,2p)$ reaction with the $S$-shell protons of
nuclei at 1 GeV are plotted versus the $S$-shell mean binding energy $E_s$.
For all nuclei (excluding $^4$He),
the effective mean nuclear density, $\bar\rho$ \cite{Hatanaka1997}, in units of
the saturation density ($\rho_0$=0.18 fm$^{-3}$) is given. The points at
$E_s$=0 correspond to the elastic proton-proton scattering
($\Theta_1$=26.0$^\circ$). The dashed curve and the solid curve
are the results of calculation in the PWIA  and the DWIA, respectively, with the
 free $NN$ interaction \cite{Chant1983}. The dotted curve corresponds to the DWIA
calculation with the relativistic effect taken into account \cite{Horowitz1986}.}
\end{figure}
For all nuclei (excluding $^4$He),
the effective mean nuclear density $\bar\rho$, normalized on the saturation nuclear
density $\rho_0\approx$~0.18 fm$^{-3}$, is given.
The experimental data were obtained in the kinematical conditions when
the nuclear $S$-shell proton before the
interaction had momentum $K$ close to zero. In this case the momentum $q$=$K_2$, transferred to a nucleus,
depended weakly on a type of the nuclear target. The actual calculation of the effective mean
nuclear density $\bar\rho$, which is determined by absorptions of the incident
and both outgoing protons, was carried out following a procedure
\cite{Hatanaka1997} using the computer code THREEDEE \cite{Chant1983}.
The potential model of a nucleus employed by the code is not  correct for
the $^4$He nucleus. The calculated value of the $\bar\rho$ in this case
is strongly unreliable \cite{Miklukho2007}. The $^4$He data should excluded
in comparing with theoretical models which differ from the PWIA. 

The points
($\circ$) and ($\bullet$) in the figure correspond to the
polarization $P_1$ and $P_2$ of the forward scattered protons at the
angle $\Theta_1$=21.0$^\circ\div$25.08$^\circ$ (with
energy $T_1$=745$\div$735 MeV) and the recoil protons
scattered at the angle $\Theta_2\simeq$ 53.2$^\circ$ (with
energy $T_2$=205$\div$255 MeV). The points at the $E_s$=0
are the polarizations $P_1$ and $P_2$
in the elastic proton-proton scattering at the angles $\Theta_1$=26.0$^\circ$
and $\Theta_2$=53.2$^\circ$ ($\Theta_{cm}$=62.25$^\circ$).
Note, that these $pp$ data were obtained by a
renormalization of the polarimeter analyzing power requiring that
the measured polarization should match the value ($P_1$=$P_2$=0.326)
 given by the current phase-shift analysis SP07 \cite{Arndt2007}.
The normalization coefficient was less than 1.06  for both polarimeters.
This correction of the analyzing power was also made  for the
polarization  data obtained in  the $(p,2p)$ experiment with nuclei.

In Fig.~4, the experimental data are compared with the results of the
non-relativistic PWIA and DWIA calculations employing
 an on-shell factorized approximation.
The final energy prescription
was used for the calculations \cite{Oers1982}.
 The dashed and solid curves correspond to the PWIA and DWIA
calculations made using the
computer code THREEDEE \cite{Chant1983}. A global optical potential \cite{Cooper1993},
parametrized in the relativistic framework and converted to the
Shr\"{o}dinger-equivalent form, was used to calculate the
distorted waves of incident and outgoing protons in the case of
DWIA. A conventional well-depth method was used to construct
bound-state wave function. To calculate the secondary proton polarization,
the THREEDEE code uses the
1986 Arndt $NN$ phase-shift analysis (SP86) \cite{Lehar1978}.
The results of the
calculations presented in Fig.~4 were normalized to a ratio of the
PWIA predictions obtained with the current phase-shift analysis
SP07 \cite{Arndt2007} and the old one SP86. The value of the ratio P(SP07)/P(SP86) was
about of 1.025.

Because the difference between the $P_1$ and $P_2$ values in the DWIA
calculations was found to be small, no more than 0.01, only the
$P_1$ values obtained from the DWIA are plotted in Fig.~4. As seen
in the figure, the difference between the PWIA and DWIA results
is quite small. This result suggests that the distortion in the
conventional non-relativistic framework does not play any
essential role in the polarization for the employed kinematic conditions
under consideration.
 The strong positive slope of the
polarizations predicted by these calculations is caused
by the kinematic effects of the binding energy of the struck
proton.
As seen in Fig.~4, there is a difference between the polarization $P_1$  calculated
in the PWIA  and  that  measured in the $(p,2p)$ reaction with all nuclei under investigation.

In Fig.~4, the experimental data (excluding the $^4$He data) are also compared with the theoretical
calculations for the case when a relativistic effect, the distortion of
the nucleon spinor, is taken into account (the $M^*_N$ curve). The calculations were
carried out in the Shr\"{o}dinger equivalent form \cite{Andreev2004} using the
THREEDEE code \cite{Chant1983}. More specifically, these calculations consist
of a non-relativistic DWIA calculations with the nucleon-nucleon
t-matrix, that is modified in the nuclear potential following a
procedure similar to that proposed by Horowits and Iqbal \cite{Horowitz1986}. In
this approach a modified $NN$ interaction in medium is
assumed due to the effective nucleon mass (smaller than the free
mass) which affects the Dirac spinors used in the calculations of
the $NN$ scattering matrix. A linear dependence of the effective
mass of nucleons on the nuclear density was assumed in the
calculations. As seen in the figure, this relativistic approach
gives the results (the dotted curve) close to the experimental values of
the forward  scattered proton polarization $P_1$
at the large transfered momenta $q$=3.2$\div$3.7 fm$^{-1}$
(see Table~4, Appendix).
This indicates that the difference
between the measured polarization $P_1$ and that calculated in the PWIA
for the nuclei $^{40}$Ca, $^6$Li, $^{12}$C, $^{28}$Si is related to
the value of the effective mean nuclear density $\bar\rho$. This
observation provides  an evidence for existing a nuclear medium effect.
 \begin{figure}
\centering\epsfig{file=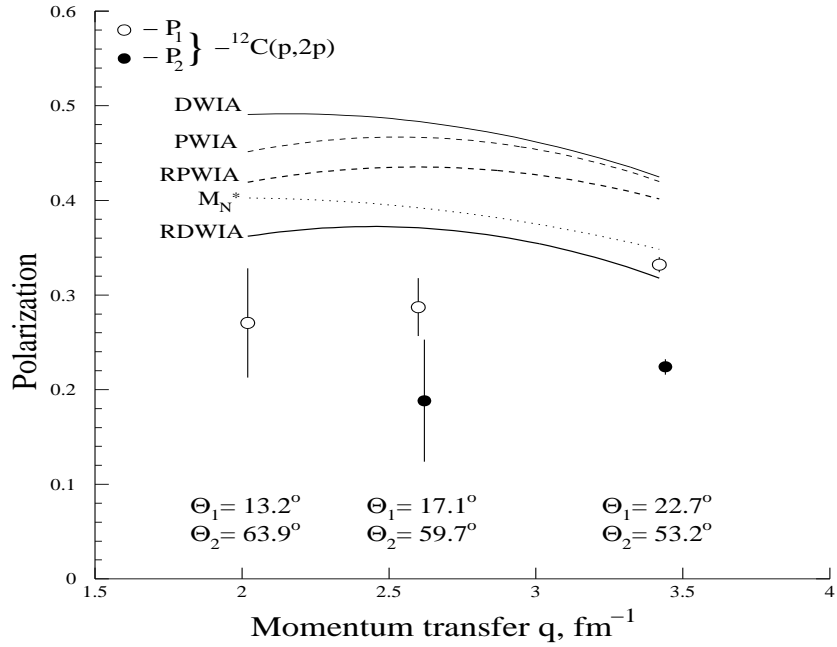,width=.8\textwidth,height=85mm}
\caption{\small Angular dependence of the polarization for the $(p,2p)$
 reaction with 1$S$-shell proton of the $^{12}$C nucleus. $P_1$ and
$P_2$ are the polarizations of forward and recoil outgoing protons, respectively.
The data are plotted as functions of the
momentum $q$ transferred to the nucleus. The thick dashed and solid curves
are, respectively, the results of the full relativistic PWIA and DWIA
calculations \cite{Hillhouse2006}.
Identifications of the other curves are the same as those in Fig.~4.
The data at $q$ $<$ 3.4 fm$^{-1}$ are taken from \cite{Andreev2004}.}
\end{figure}
The Fig.~5 demonstrates that the relativistic approach (the dotted curve), mentined above,
failed to reproduce the experimental polarization data for the $^{12}$C nucleus
at the transferred momentum $q$ less than 3.4 fm$^{-1}$.
The thick solid curve in the
figure corresponds to the calculations in a fully relativistic model
based on the distorted wave impulse approximation (RDWIA). In this model the
four-component scattering and bound-state wave functions are governed by
the relativistic Dirac equation \cite{Hillhouse2006}. Note, that in the
full relativistic approach both real and imaginary components of the
spin-orbit potential, which are crucial for describing polarization
phenomena, are directly related to the Lorentz properties of
mesons mediating the strong nuclear force. This microscopic connection
does not exist within the framework of the non-relativistic Shr\"{o}dinger
equation. In this case the spin-orbit interaction is usually introduced by hand
and the interaction parameters are fitted to provide a good phenomenological
description of the elastic scattering data.
As seen from the figure, this relativistic approach also overestimates
the experimental values of the $P_1$ polarization
at the transferred momenta $q$ $<$  3.4 fm$^{-1}$.
The difference may be related to
another possible medium effect due to modifications of the exchanged
meson masses and meson-nucleon coupling constants in the $NN$ interaction.
Krein et al. have shown in the relativistic Love-Franey
model that these modifications cause significant
changes in the spin observables
\cite{Krein1995}. A such type of modification was investigated
in \cite{Hillhouse2006} using
our early obtained experimental data on the polarization in the $(p,2p)$ reaction with
the $^{12}$C nucleus measured in a wide
range of the momentum transfer $q$ \cite{Andreev2004}. Note,
we essentially improved a statistical accuracy of the polarization
measurement at  the $q$=3.4 fm$^{-1}$.

As seen in Fig.~4, there is a systematic difference between
the $P_1$ and $P_2$ values, though they have the same values in
the case of the elastic $pp$ scattering.
The Horowits and Iqbal relativistic approach \cite{Horowitz1986}, considered above,
also gives practically equal values of the polarizations.
In Fig.~6, the relative difference $g$ = ($P_2$ - $P_1)$/$P_1$
\begin{figure}
\centering\epsfig{file=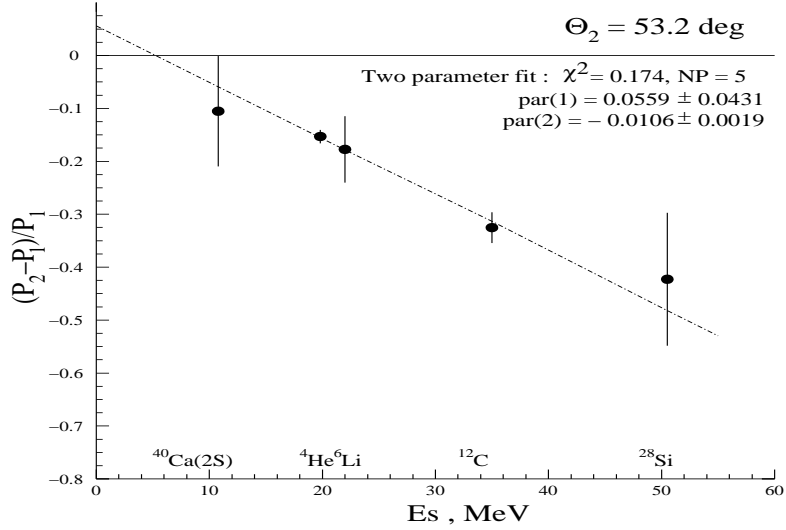,width=.75\textwidth,height=70mm}
\caption{\small Relative difference $g$=($P_2$-$P_1$)/$P_1$ of the measured polarizations $P_2$, $P_1$
for all investigated nuclei.
The dash-dotted curve is a linear fit to the experimental data.}
\end{figure}
between the experimental values of the
polarizations for all investigated nuclei
is plotted versus the S-shell proton mean binding energy $E_s$.
The dash-dotted line in the figure is the linear fit to the experimental data.

An origin of the stronger reduction of the polarization $P_2$ may be related to
a manifistation of the recoil proton interaction with the short-lived two-nucleon or
three-nucleon dense associations arose due to a fluctuation of the nucleon density
in nuclear matter \cite{Blokhintsev1957,CLAS2006}. The recoil S-shell proton with the momentum $K_2$=$q$
($T_2 \sim $200 MeV) and with the polarization $P_2 \simeq P_1$ elastically collides
with the dense nucleon association. As a result, the S-shell proton of the association leaves
a nucleus with the same momentum and, according to Pauli exclusion principle,
with the opposite sign of the polarization $P_2 \simeq - P_1$. Taking into account a contribution from
this depolarization mechanism, a probability of which is denoted as $\alpha$,
the averaged polarization of
the recoil proton $\bar{P_2}$  is
\begin{equation}
\label{equation:p2}
\bar{P_2} = (1-\alpha)P_1 + \alpha(-P_1) = P_1(1 - 2\alpha)\\,
\end{equation}
and then the relative difference $g$ between the polarizations $\bar{P_2}$ and $P_1$ is
\begin{equation}
\label{equation:g}
g = \frac{(\bar{P_2} - P_1)}{P_1} = - 2 \alpha\\.
\end{equation}
The value of $g$ extracted from the experimental data for
the $^4$He and $^{12}$C nuclei (see Table~4 in Appendix) is equal to $g$ = - 0.153 $\pm$ 0.013
and $g$ = - 0.325 $\pm$ 0.029, respectively.

The spin correlation parameters $C_{ij}$ in
the reactions with the $^4$He and
$^{12}$C nuclei were for the first time measured using an
unpolarized 1 GeV proton beam.
The results of the $C_{ij}$ measurements
are given in Fig.~7 (Table~5 in Appendix).
\begin{figure}
\centering\epsfig{file=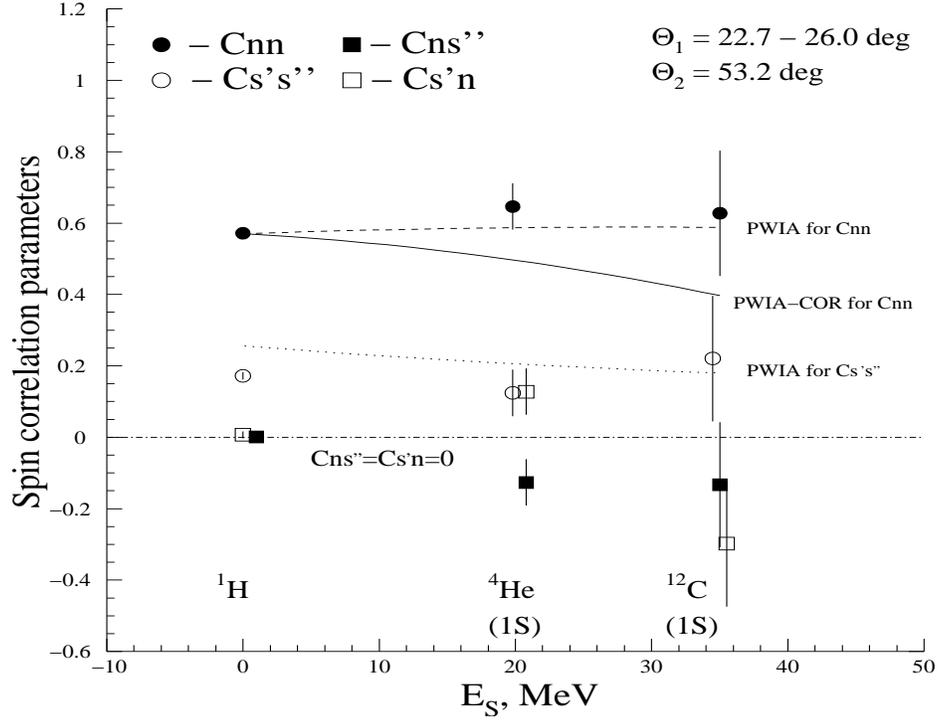,width=.9\textwidth,height=95mm}
\caption{\small Spin correlation parameters $C_{ij}$ in the $(p,2p)$ reaction at 1 GeV
with the $S$-shell protons of the $^4$He and $^{12}$C nuclei at the
secondary proton scattering angles $\Theta_2$=53.22$^\circ$,
$\Theta_1$=24.21$^\circ$ and $\Theta_2$=53.22$^\circ$, $\Theta_1$=22.71$^\circ$,
respectively. The points at $E_s$=0 correspond to the averaged values of
the polarization parameters in the
elastic proton-proton scattering ($\Theta_1$=26.0$^\circ$,
$\Theta_{cm}$=62.25$^\circ$), obtained in 2008-2009 years
(Table~6 in Appenix). The dashed  and  dotted curves are the results
of the PWIA calculation of the $C_{nn}$ and $C_{s^,{s^{,,}}}$ spin
correlation parameters. The solid curve corresponds to
the result of the PWIA-COR calculation
taking into account the depolarization mechanism described in the text.}
\end{figure}

In Table~5, the measured
mean values of the $C_{ij}$ for the accidental coincidence
background obtained in investigating the $(p,2p)$ reaction with
nuclei $^4$He, $^{12}$C and $^{28}$Si are also presented.
In Fig.~7, the dashed and dotted curves correspond to the PWIA
calculations for the $C_{nn}$ and $C_{s^,{s^{,,}}}$.
In these calculations, the current Arndt
phase-shift analysis (SP07) was used \cite{Arndt2007}. The  $C_{s^,{s^{,,}}}$
parameter was found by taking into account its distortion in the
magnetic fields of the MAP and NES spectrometers due to an
anomalous proton magnetic moment \cite{Lock1973}. The points at the mean
binding energy value $E_s$=0 correspond to the
elastic proton-proton scattering (see Table~6 in Appendix). These points
in the figure are the averaging result
of the $pp$ experimental data obtained in 2008 and 2009 years when  the
nuclei $^4$He and $^{12}$C were investigated.

As seen in Fig.~7, the differences between the measured values of the $C_{nn}$
and those calculated in the PWIA are within the statistical errors.
The measured value of $C_{s^,{s^{,,}}}$
in the elastic proton-proton scattering
strongly differs from the SP07 prediction. This can be explained
as a lack of the spin correlation
parameter data from the elastic $pp$ scattering experiments
in the base of the current phase-shift analysis.
 Due to the parity conservation in the  elastic proton-proton
scattering, the spin correlation parameters $C_{ns''}$ and
$C_{s'n}$ should be equal to 0. In Fig.~7, this is confirmed by the experimental
data at the $E_s$=0. For a $(p,2p)$ reaction with
nuclei the parity in the $pp$ system can be violated since
there exists a residual nucleus in the knockout
process. However, in this case, according to Pauli exclusion principle,
a relation  between the parameters  in the
center-of-mass $pp$ interaction system $C_{ns''}$= - $C_{s'n}$
should be valid \cite{Faisner1959}.
This relation can be violated in the laboratory
system because of a distortion of the parameters
by the magnetic fields of both spectrometers.

In Fig.~7, the experimental $C_{nn}$ data for the $^4$He and $^{12}$C nuclei
are also analysed in the framework of the plane
wave impulse approximation with the correction to a contribution from
the depolarisation mechanism  PWIA-COR related to
the mentioned above recoil-proton elastic interaction with the dense nucleon associations
in a nucleus (the solid curve).
 The averaged value of the spin correlation parameter $\bar C_{nn}$ in the PWIA-COR model
was calculated  using equation similar to Eq.~(\ref{equation:p2})
for the reoil proton polarization $\bar P_2$:
\begin{equation}
\label{equation:cnn-cor}
\bar C_{nn} = (1 - \alpha)C_{nn} + \alpha (- C_{nn}) = C_{nn} (1 - 2\alpha)  = C_{nn}(1 + g)\\,
\end{equation}
where the $C_{nn}$ is the result of the PWIA calculation (the dashed curve) and the
$g<$0 is the relative difference of the $P_2$ and $P_1$ experimental polarization
values (Eq.~(\ref{equation:g})), calculated  above for the $^4$He and $^{12}$C nuclei.
As seen in Fig.~7, the $C_{nn}$ experimental values for the nuclei
exceed those calculated in the PWIA-COR.
Therefore, the nuclear medium increases the $C_{nn}$ parameter value in the $(p,2p)$ reaction
in comparison with the elastic $pp$ scattering. On the contrary, as shown above,
the $NN$-scattering amplitudes are modified in nuclear matter so that the
polarization in the reaction is less than that in the elastic $pp$ interaction.

\section{Summary}
~~~ The polarizations $P_1$ and $P_2$ of both
secondary protons from the $(p,2p)$ reaction with the 1$S$-shell
protons of the nuclei $^4$He, $^6$Li, $^{12}$C, $^{28}$Si and with
2$S$-shell protons of the $^{40}$Ca nucleus were measured at 1 GeV
energy of unpolarized proton beam. The spin correlation parameters
$C_{ij}$ in the reaction with the nuclei $^4$He  and $^{12}$C were
for the first time obtained. The experiments were performed at the
non-symmetric scattering angles of the final state protons
$\Theta_1$=21$^\circ\div$25$^\circ$  and $\Theta_2\simeq$
53.2$^\circ$ in the coplanar quasi-free scattering geometry with a
complete reconstruction of the reaction kinematics. The
measurements were done in the kinematical conditions when the
nuclear $S$-shell proton (before the interaction) had the momentum
$K$ close to zero. In this case the transferred momentum $q$,
which is equal to the recoil proton momentum $K_2$, depended
weakly on a type of the nuclear target ($q$=3.2$\div$3.7
fm$^{-1}$).

The observed reduction of the polarization $P_1$ of the forward scattered protons
in the reaction with the nuclei $^6$Li, $^{12}$C, $^{28}$Si and $^{40}$Ca
in comparison with that calculated in the framework of the
PWIA and DWIA
depends on the effective mean nuclear density.
The calculations in the DWIA, in the framework of which
a modification of nucleon Dirac spinor is taken into account,
reproduce the magnitude of the reduction well.
The inessential difference between the non-relativistic PWIA and DWIA
 calculations points out to a small
depolarization of the secondary protons because of the proton-nucleon rescatterings in a nucleus.
All these facts indicate a modification of the proton-proton scattering matrix
in the nuclear medium.

The experiment with the $^4$He nucleus shows that the magnitude
of the $P_1$ polarization reduction in the light nuclei at least (A$< $12) is also
determined by the mean binding energy of the $S$-shell proton of a nucleus.
Another significance of the experiment with the $^4$He nucleus
is the possibility to see the medium effect without
any contribution from the multi-step processes. These processes can
take place when there are nucleons of outer shells as in other nuclei.
The experiment also points out that the multi-step processes are not the main origin of the observed
negative difference of the polarizations of the recoil proton $P_2$ and  the
forward scattered proton $P_1$.

The systematic relative difference between the $P_2$ and $P_1$ polarizations was
observed in investigating the $(p,2p)$ reaction with nuclei. This effect depends likely
 on the value of the mean binding energy $E_s$ of the $S$-shell protons.
 An origin of the
effect
may be related to the depolarization mechanism of
the recoil proton interaction with the short-lived dense nucleon associations
in a nucleus.
This depolarization mechanism  decreases the value
of the spin correlation parameter $C_{nn}$ in comparison with that calculated
in the PWIA. In this reason, the $C_{nn}$ experimental values
for the nuclei $^4$He and $^{12}$C
exceed those calculated in the PWIA with a contribution from the mechanism included.
Therefore, the nuclear medium affects
the $C_{nn}$ parameter in the $(p,2p)$ reaction increasing its value
in comparison with that in the elastic $pp$ scattering. On the contrary,
the $NN$ scattering amplitudes are modified in nuclear matter so that the
polarization in the $(p,2p)$ reaction is less than that in the elastic $pp$ interaction.
\newpage
\section{Acknowledgments}
~~~ This work is partly supported by the
Grant of President of the Russian Federation for Scientific
School, Grant-3383.2010.2.

The authors are grateful to PNPI 1~GeV proton accelerator staff
for stable beam operation. We thank members of PNPI HEP
Radio-electronics Laboratory for providing the CROS-3 proportional
chamber readout system.

Also, the authors would like to express their gratitude to
A.A.~Vorobyov and S.L.~Belostotski for their support and fruitful
discussions.

\section{Appendix: Experimental results}
\begin{table}[h]
\caption{Secondary proton polarizations $P_1$ and $P_2$
in the $(p,2p)$ reaction at 1 GeV with the $S$-shell protons of nucleus at lab.
angles $\Theta_1$ and $\Theta_2$}
\label{table:pol}
\begin{center}
\begin{tabular}{c|c|c|c|c|c|c|c|c}
\hline
Nucleus & $\Theta_1$ & $\Theta_2$ & $T_1$ & $T_2$ & $ q $ & $P_1$ & $P_2$  &
$\bar\rho/\rho_0$ \\
     &deg. & deg. & MeV & MeV & fm$^{-1}$ &       &       &                   \\
\hline
$^4$He (1S) & 24.21 & 53.22 & 738 & 242 & 3.6 & .328$\pm$.003 & .278$\pm$.003 &  
\\\hline
$^6$Li (1S) & 24.0 & 53.25 & 739 & 239 & 3.6 & .310$\pm$.015 & .255$\pm$.015 & .19 
\\
\hline
$^{12}$C (1S) & 22.71 & 53.22 & 746 & 219 & 3.4 & .332$\pm$.008 & .224$\pm$.008 & .31 
\\
\hline
$^{28}$Si (1S) & 21.0 & 53.22 & 746 & 204 & 3.2 & .350$\pm$.038 & .202$\pm$.038 & .30 
\\
\hline
$^{40}$Ca (2S) & 25.08 & 53.15 & 734 & 255 & 3.7 & .311$\pm$.025 & .285$\pm$.023 & .07 
\\
\hline
\end{tabular}
\end{center}
\end{table}

\begin{table}\caption{Spin correlation parameters $C_{ij}$ in the $(p,2p)$
reaction at 1 GeV with the $S$-shell protons of the $^4$He and
$^{12}$C nuclei at lab.~angles $\Theta_1$  and
$\Theta_2$=53.22$^\circ$. The line "Background" corresponds to the
measured mean values of the $C_{ij}$ for the accidental
coincidence background, obtained in investigating the reaction
with the nuclei $^4$He, $^{12}$C and $^{28}$Si} \label{table:cnn}
\begin{center}
\begin{tabular}{c|c|c|c|c|c}
\hline
Nucleus & $\Theta_1$ & C$_{nn}$   & C$_{s^,}$$_{s^{,,}}$ & C$_{ns^{,,}}$ & C$_{s^,}$$_n$
\\
        & deg.       &          &          &          &
\\
\hline
$^4$He  & 24.21 & .647$\pm$.065 & .124$\pm$.065 & -.126$\pm$.065 & .128$\pm$.065
\\
\hline
$^{12}$C & 22.71 & .628$\pm$.176 & .220$\pm$.176 & -.133$\pm$.176 & -.298$\pm$.176
\\
\hline
Background&      & -.006$\pm$.019 & .01$\pm$.019  & .016$\pm$.019  & .001$\pm$.019
\\
\hline
\end{tabular}
\end{center}
\end{table}

\begin{table}
\caption{ Spin
correlation parameters C$_{ij}$ in the elastic proton-proton
scattering at 1 GeV at lab. angles $\Theta_1$=26.0$^\circ$   and
$\Theta_2$=53.22$^\circ$ ($\Theta_{cm}$=62.25$^\circ$).
Results of 2008-2010 years. The current phase-shift analysis
predicts the $C_{nn}$ value of 0.57. The statistical errors of
the polarization $P_1$  and
$P_2$ measurements are given. The systematic uncertainty in the polarization measurements
was about of  $\pm$ 0.004 and  $\pm$ 0.0025 for the MAP and NES polarimeters, respectively
\cite{Miklukho2010}.}
\label{table:cnn-pp}
\begin{center}
\begin{tabular}{c|c|c|c|c|c|c}
\hline
Year &  C$_{nn}$  & C$_{{s^,}{s^{,,}}} $ & C$_{ns^{,,}}$ & C$_{{s^,}n}$ & $\delta$P$_1$ & $\delta P_2$
\\
\hline
2008 & .557$\pm$.013 & .185$\pm$.013 &  .010$\pm$.013 & -.003$\pm$.013 & .0008 & .0007
\\
\hline
2009 & .589$\pm$.015 & .154$\pm$.015 & -.012$\pm$.015 & .020$\pm$.015 & .0011  & .0007
\\
\hline
2010 & .575$\pm$.030 & .176$\pm$.030 & -.029$\pm$.030 & -.003$\pm$.030 & .0021  & .0015
\\
\hline
\end{tabular}
\end{center}
\end{table}

\end{document}